\documentclass[letter]{aa}
\usepackage{txfonts,graphicx,natbib}

\begin{document}

  \title{First detection of photospheric depletion in the LMC\thanks{based
  on observations collected at the European Southern Observatory, Chile
  (programme 074.D-0619(A))}}

   \author{M. Reyniers\thanks{Postdoctoral fellow of the Fund for
          Scientific Research, Flanders}
          \and
          H. Van Winckel
          }
   \offprints{M. Reyniers}

   \institute{Instituut voor Sterrenkunde, Departement Natuurkunde en
              Sterrenkunde, K.U.Leuven, Celestijnenlaan 200D, 3001 Leuven,
              Belgium
              \email{maarten@ster.kuleuven.be}}

   \date{Received December 5, 2006; accepted December 31, 2006}

\abstract
  {Recent photospheric abundance studies of galactic field RV Tauri stars show
  that depletion of refractory elements is rather common in these evolved
  objects.}
  {The process that creates this chemical anomaly is not understood well, but
  it probably requires the presence of gravitationally bound dust in a binary
  system. We test for the presence of depletion in extra-galactic objects.}
  {A detailed photospheric abundance study on the basis of high-quality
  UVES spectra was performed on the RV Tauri star in the LMC:
  MACHO\,82.8405.15.  
  Abundances were derived using a critically compiled line list with
  accurate log(gf) values and the latest Kurucz model atmospheres.}
  {With [Fe/H]\,=\,$-$2.6 in combination with [Zn/Fe]\,=\,$+$2.3 and
  [S/Ti]\,=\,$+$2.5, MACHO\,82.8405.15 displays a strong depletion abundance
  pattern.  The effect of the depletion is comparable to the strongest
  depletions seen in field Galactic RV\,Tauri stars.}
  {The chemical analysis of MACHO\,82.8405.15 proves that the depletion process
  also occurs in the extragalactic members of the RV Tauri pulsation class. Our
  program star is a member of a larger sample of LMC RV Tauri objects. This
  sample is unique, since the distances of the members are well-constrained.
  Further studies of this sample are therefore expected to gain deeper insight
  into the poorly understood depletion phenomenon and of the evolutionary status
  of RV\,Tauri stars in general.
}


   \maketitle

\section{Introduction}
It is now almost one century ago that \citet{seares08} noted how RV Tauri,
the prototype of the RV Tauri variables, shows a light curve with subsequent
deep and shallow minima. This observation, together with the required limits
on the spectral type (F to K),  is still used as the defining characteristic
for the RV Tauri variables.
\citet{preston63} defined three spectroscopic classes: The stars of type RVA
have spectral type G-K, and show strong absorption lines but normal CN or CH
bands. 
The RVB stars are generally somewhat hotter (spectral type F), weak-lined
objects that show enhanced CN and CH bands. 
RVC stars are also weak-lined stars, but they show normal bands of CH and CN.

The exact evolutionary stage of RV Tauri stars is still not very clear, but
they are probably in the post-AGB stage of evolution, as first argued by
\citet{jura86}. Many of the RV Tauri stars show an infrared excess, which is
caused by the circumstellar dust that was likely formed by heavy mass loss during
the preceding AGB phase. The MACHO experiment in the 90s revealed the first
{\em extragalactic} RV Tauri stars in the Large Magellanic Cloud.
\citet{alcock98} showed the light curves of about a dozen RV Tauri stars in the
LMC, with ``formal periods'' (the time between two successive minima) between
40 and 120 days, defined a single PL relation for these stars, and strengthened
the post-AGB interpretation of RV Tauri stars by their luminosities. 

The extensive chemical studies of the photospheres of field RV Tauri stars
showed, however, that none of the RV Tauri stars shows evidence of post 3rd
dredge-up chemical enrichments, such as enhanced carbon or enhanced s-process
abundances \citep[except for maybe V453\,Oph, see][]{deroo05}. Instead, the
abundance studies
\citep[see e.g.][and references therein]{giridhar00, giridhar05, maas05}
reveal that severe abundance anomalies are observed in mainly RVB stars, with 
underabundances of elements with a high condensation temperature. The origin of
these {\em depletion patterns} probably involves chemical fractionation due to
dust formation in the circumstellar environment, followed by a decoupling of
the gas and dust with a reaccretion of the cleaned gas on the stellar
photosphere. It was realised that the spectroscopic classes of
\citeauthor{preston63} could be attributed to a different degree of depletion,
with the RVB stars the most heavily depleted. The whole process probably
requires a stable circumstellar environment, like a stable disk
\citep{waters92}. Since a stable disk is only likely to form in a binary system,
it is believed that all depleted RV Tau stars are probably binaries
\citep{vanwinckel03, deruyter06}. Note that depletion does not seem to occur in
globular cluster RV Tauri stars, or in field RV\,Tauri stars of low
metallicity ([Fe/H]\,$<$\,$-$1). Recently, we \citep{reyniers06} reported on
the intrinsic metal-poor RV\,Tauri star MACHO\,47.2496.8 in the LMC, which is
unique for displaying a very strong s-process enrichment. This corroborates
the post-AGB nature of this object, but adds to the chemical diversity observed
in this pulsation class.

Preliminary work by \citet{lloydevans04} showed, on the basis of low-resolution
spectra, that at least two RV Tauri stars found through the MACHO experiment
are RVB stars, and hence potentially depleted. In this letter, we focus on the
brightest of both objects: MACHO\,82.8405.15 (m$_V$\,$=$\,14.3).
\citet{alcock98} found a period of 93.1\,days for this object and a mean
absolute visual magnitude of M$_V$\,$=$\,$-$4.52. The amplitude of the
pulsation is significant, with a peak-to-peak of 0.89 in the MACHO V magnitude.
Here, we present a detailed abundance analysis of MACHO\,82.8405.15, based on
high-resolution, high signal-to-noise VLT-UVES spectra.

\section{Observations and analysis}
\begin{figure}
\resizebox*{\hsize}{!}{\includegraphics{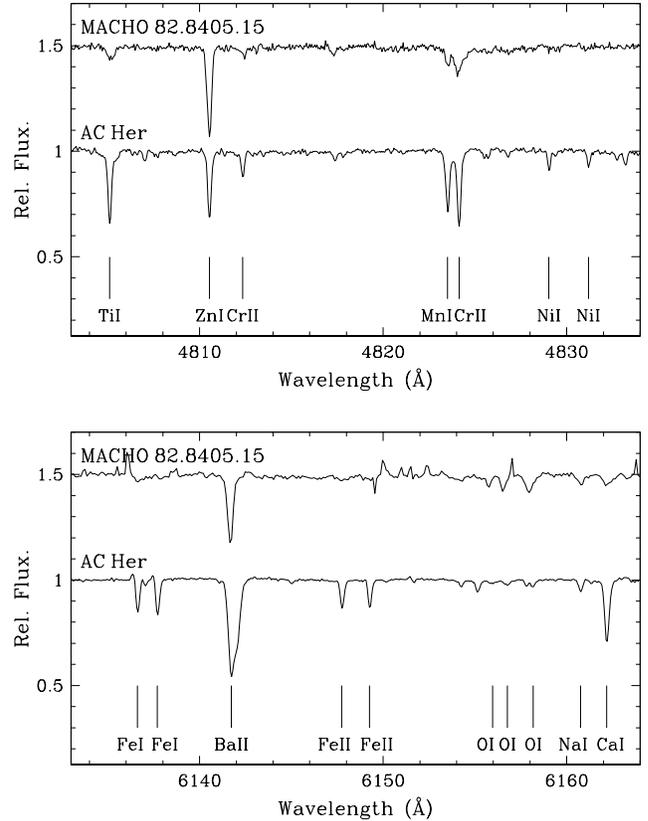}}
\caption{The VLT+UVES spectrum of MACHO\,82.8405.15, together with the spectrum
of AC\,Her, around the 4810\,\AA\ Zn line {\em (upper panel)} and the
6155\,\AA\ oxygen triplet {\em (lower panel)}. AC\,Her is a depleted field RV
Tauri star that is somewhat cooler (T$_{\rm eff}$\,$=$\,5500\,K) than
MACHO\,82.8405.15. It is extensively discussed in \citet{vanwinckel98}. In the
upper panel, the Zn line is remarkably strong, while lines of other iron peak
elements are almost absent. In the lower panel, the oxygen triplet around
6155\,\AA\ is clearly visible, indicating the hotter temperature for
MACHO\,82.8405.15. Note that the spectrum in the lower panel has quite a lot of
telluric emission. Correction for this emission is difficult, since this part
of the spectrum is observed with the image slicer, hence no sky window is
available.}\label{fig:samplespctrm}
\end{figure}

Spectra were taken with UVES on the VLT-UT2 (Kueyen) in visitor mode by
one of us (HVW). Details of the observations, together with some indicative
signal-to-noise (S/N) ratios, can be found in Table \ref{tab:observationsdtls}.
The spectra were taken just before minimum light.
The reduction was performed in the dedicated ``UVES context'' of the ESO MIDAS
software packet, including all standard reduction steps. Optimal extraction was
used, except for the spectrum taken in the first night, since this was taken
with the slicer and optimal extraction is not implemented for this type of
data. A sample spectrum can be found in Fig. \ref{fig:samplespctrm}.

\begin{table}\caption{Log of the high-resolution VLT-UVES observations, with
small spectral gaps between 5757\,\AA\ and 5833\,\AA\ and between 8521\,\AA\ and
8660\,\AA\ due to the spatial gap between the two UVES CCDs.}\label{tab:observationsdtls}
\begin{center}
\begin{tabular}{rrrrr}
\hline\hline
 date & UT    & exp.time & wavelength & S/N  \\
      & start &  (sec)   & interval (\AA) &     \\
\hline
 2005-02-08 & 01:43 & 3600 & 4780$-$6808 & 130 \\ 
            &       &      & 3758$-$4983 & 110 \\ 
\raisebox{1.5ex}[0pt]{2005-02-09}&\raisebox{1.5ex}[0pt]{02:37}& \raisebox{1.5ex}[0pt]{7200} &\raisebox{1.5ex}[0pt]{\Big\lbrace} 6705$-$10084 & 120 \\ 
\hline
\end{tabular}
\end{center}
\end{table}


The method of analysis is similar to the method used in our previous
publications on similar stars \citep[e.g.][]{deroo05}. First, a complete
line identification was performed, using the line lists of \citet{thevenin89,
thevenin90}. Then, lines with a clean profile and an equivalent width
smaller than 150\,m\AA\ were selected for abundance calculation purposes. The
log(gf) values of the selected lines were taken from the line list with
accurate oscillator strengths described in \citet{vanwinckel00}. Lines without
an accurate log(gf) value were discarded. For the abundance calculation, we
used the latest Kurucz models ({\tt\small http://kurucz.harvard.edu}), in
combination with the LTE line analysis program MOOG by C. Sneden (April 2002
version). Since initially only 24 Fe\,{\sc i} and 3 Fe\,{\sc ii} lines were
found with our usual line list, 11 Fe\,{\sc i} and 4 Fe\,{\sc ii} lines were
added from \citet{thevenin89,thevenin90}, although these papers are generally 
regarded as a less accurate source for oscillator strengths.

In our search for the atmospheric parameters of MACHO\,82.8405.15, we realised
that the classical ``spectroscopic'' method of studying the iron lines failed.
The usual technique of demanding excitation equilibrium for the Fe\,{\sc i}
lines resulted in an effective temperature of T$_{\rm eff}$\,$=$\,7000\,K,
but such a high effective temperature gives unrealistic supersolar zinc (Zn)
abundances. Moreover, at T$_{\rm eff}$\,$=$\,7000\,K, ionisation equilibrium
can only be attained with a high gravity of $\log g$\,$=$\,2.5. The reason
the classical method fails is not clear, but is probably related to the
dynamical character of the photosphere (see Sect.~\ref{sect:radialvlcts})
and/or to the well-known non-LTE effect of UV overionisation in the atmospheres
of late-type stars \citep[e.g.][]{shchukina05}. This effect is expected to be
greater in metal-poor stars as a result of the lower electron density and of
the weaker UV blanketing.

\begin{figure}
\resizebox*{\hsize}{!}{\rotatebox{-90}{\includegraphics{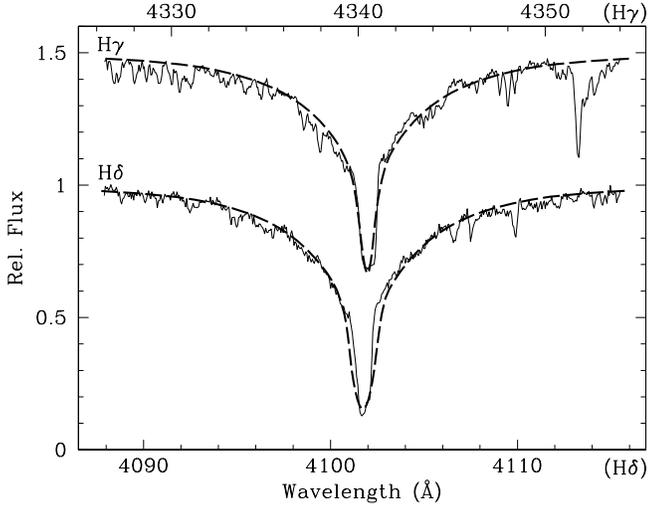}}}
\caption{The final  H$\gamma$ and H$\delta$ Balmer fits.}\label{fig:balmerfts}
\end{figure}

\begin{figure}
\resizebox*{\hsize}{!}{\includegraphics{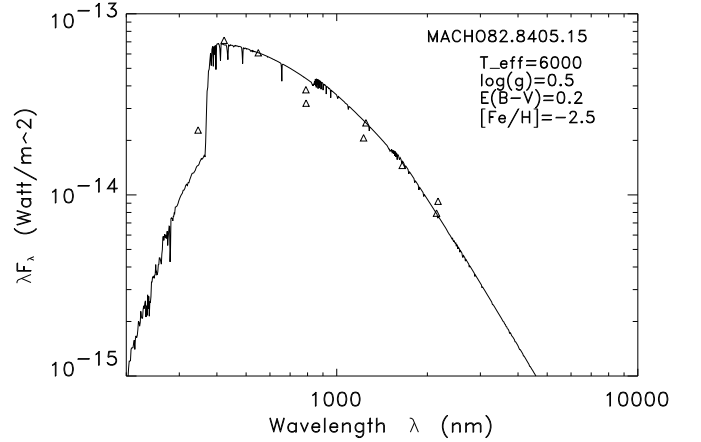}}
\caption{The spectral
energy distribution (SED) of MACHO\,82.8405.15. The dereddened photometry is
shown with triangles, while the Kurucz model is overplotted with a solid line.
The photometry includes: U, B, V magnitudes in the Geneva system; I in the
Cousins system; I, J, and K from DENIS; and J, H, and K from 2MASS. The
optical photometry shown here is taken at a more luminous phase than the 
UVES spectroscopy on which the atmospheric parameters are based. The SED fits,
however, were only used to constrain the reddening towards MACHO\,82.8405.15.}\label{fig:machosd}
\end{figure}

We had to follow an alternative scheme to deduce the atmospheric parameters.
First, we estimated the gravity from the luminosity that is given in
\citet{alcock98}. This estimate is of course dependent on the adopted
effective temperature, the initial mass, and the reddening. However, for all temperatures
compatible with an F-star spectrum, a low gravity of  $\log g$\,$\le$\,1.0 is
found. A further decrease in the parameter space was obtained through fitting
the H$\gamma$ and H$\delta$ Balmer lines using Kurucz Balmer profiles. The
H$\alpha$ and H$\beta$ lines were not used for this purpose since they show strong
emission features. The wings and the cores of the two lines were fitted
simultaneously, with more emphasis on a good fit for the wings. We could
constrain the parameter space further to models with
T$_{\rm eff}$\,$\le$\,6000\,K by this Balmer-profile fitting.
From a comparison of photometry taken with the Swiss Euler telescope at La Silla
\citep[see][for more details on these observations]{reyniers06}
with synthetic colours, we inferred that the
reddening is certainly small, with an upper limit of E(B-V)\,$=$\,0.2.
If we combine this information with the gravity from the luminosity {\em and}
the Balmer fits, we could conclude that the only consistent temperature is
T$_{\rm eff}$\,$=$\,6000\,K. The final choice between the two consistent
gravities at this temperature, 0.5 and 1.0, was made on the basis of the
consistency in the abundances, resulting in $\log g$\,$=$\,0.5. The
 H$\gamma$ and H$\delta$ Balmer fits with the final parameters are shown in
Fig.~\ref{fig:balmerfts}; the spectral energy distribution is shown in
Fig.~\ref{fig:machosd}.
Note that there is a phase difference of 0.24 between the UVES spectra
on which the Balmer fitting was performed and the photometry, implying
a high uncertainty in the derived parameters (at least $\pm$250\,K
in T$_{\rm eff}$ and $\pm$1.0 in $\log g$). We want to stress, however, that
in the parameter determination, the photometry was used solely to constrain
the reddening towards MACHO\,82.8405.15.

\section{Radial velocities of individual lines}\label{sect:radialvlcts}
\begin{figure}
\resizebox*{\hsize}{!}{\rotatebox{-90}{\includegraphics{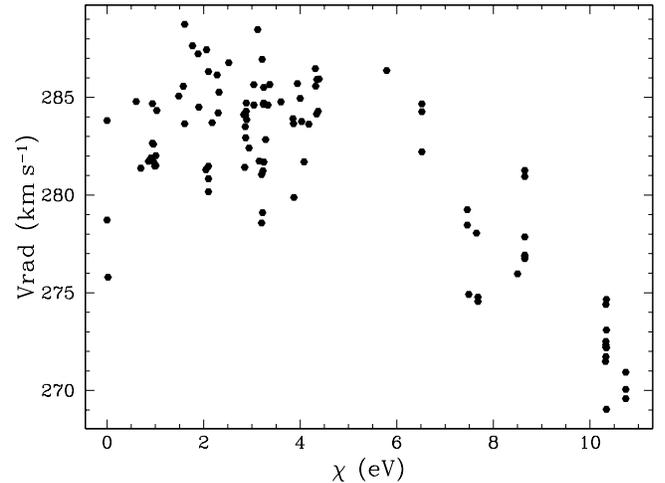}}}
\caption{The clear correlation between the heliocentric radial velocities
of individual lines and their lower excitation potentials.}\label{fig:vradtrnd}
\end{figure}

Before turning to the actual abundances, the radial velocities of the
individual spectral lines need to be noted. During our analysis, we
noticed that some lines were more velocity-shifted than others. Moreover, a
clear correlation exists between the radial velocity and the excitation
potential of a line, with lower velocities for lines with a high excitation
potential. This is illustrated in Fig. \ref{fig:vradtrnd}. It is indicative of
an optical depth effect due to a differential velocity field in the
line-forming region of the dynamical photosphere of this pulsating star. In
field RV\,Tauri stars, this shock-wave passage can lead to line-splitting at
some pulsational phases \citep[e.g.][]{gillet90}. Very similar velocity
effects were also detected in the small-amplitude pulsating post-AGB star
HD\,56126 \citep{lebre96,barthes00,fokin01}, and were again attributed to the
passage of a shock wave. Note that the UVES spectra were taken just before
minimum light, i.e. the phase at which a new shock wave is expected to develop
\citep{gillet89}. The adoption of plane-parallel, LTE Kurucz models can thus
only be a first approximation of the real photosphere. 

\section{Results and discussion}\label{sect:resultsdscssn}
The final abundances can be found in Table \ref{tab:abundancerslts}. The first
column of this table gives the actual ion; the second column gives the
number of lines used; the third one is the mean equivalent width; the fourth
column gives the absolute abundances derived
$\log\epsilon$\,$=$\,$\log$(N(el)/N(H))+12; $\sigma_{\rm ltl}$, the fifth
column, is the line-to-line scatter; the sixth column gives the abundance
relative to the sun [el/H]. For the references of the solar abundances (seventh
column) needed to
calculate the [el/H] values: see \citet{reyniers06}. The dust condensation
temperatures (last column) are taken from \citet{lodders03}. 

\begin{figure}
\resizebox*{\hsize}{!}{\rotatebox{-90}{\includegraphics{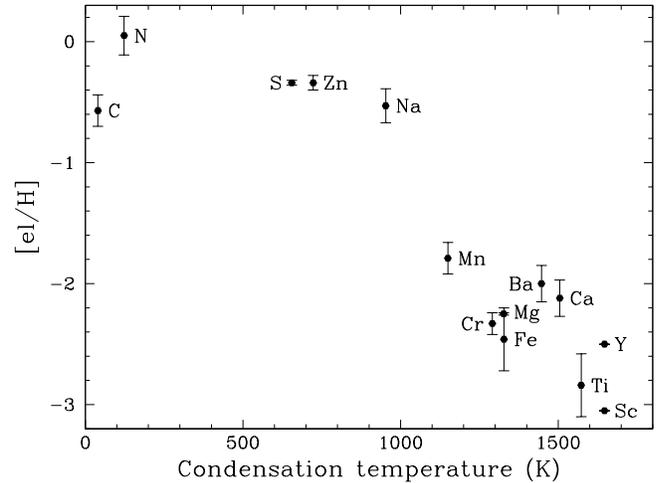}}}
\caption{The anti-correlation between the abundance of an element
([el/H]) and its condensation temperature. This indicates that a very efficient
depletion process has taken place in MACHO\,82.8405.15. The errorbar is
the line-to-line scatter listed in Table~\ref{tab:abundancerslts}.}\label{fig:abundancerslts}
\end{figure}

The abundances are
also graphically presented in Fig. \ref{fig:abundancerslts}, where the
abundance relative to solar [el/H] is plotted against condensation temperature.
The clear anti-correlation as seen on Fig. \ref{fig:abundancerslts} is
undoubtfully recognised as a depletion pattern: the slightly subsolar
abundances of Zn and S are expected for a field LMC-star, while iron is
depleted by more than a factor of 300. The CNO elements have low condensation
temperatures and are typically expected
not to be depleted. The low carbon abundance of [C/H]\,$=$\,$-$0.6 makes clear
that MACHO\,82.8405.15 has not experienced a 3rd dredge-up. One remarkable result
is the detection of two forbidden $[$O\,{\sc i}$]$ lines at 6300.23\,\AA\ and
6363.88\,\AA, both in absorption. The abundances derived from these lines are
much higher than the abundances derived from the O\,{\sc i} triplet at
6155\,\AA\ (see Table \ref{tab:abundancerslts}). The $[$O\,{\sc i}$]$ lines are
known not to be very sensitive to non-LTE effects, so this large abundance
difference is not expected. The two oxygen abundances can be brought
in agreement if one adopts a slightly lower temperature (5750\,K) combined with
a very
low gravity of $\log g$\,$=$\,$-$0.4. Such a low gravity is, however, not
compatible with the luminosity. The abundances of the s-process elements Y and
Ba show an expected behaviour 
and are probably not intrinsically enhanced.

\begin{table}
\caption{Abundance results for MACHO\,82.8405.15.}\label{tab:abundancerslts}
\begin{center}
\begin{tabular}{lrrrrrrr}
\hline\hline
\multicolumn{7}{c}{\rule[-0mm]{0mm}{5mm}{\Large\bf MACHO\,82.8405.15}}\\
\multicolumn{7}{c}{
$
\begin{array}{r@{\,=\,}l}
{\rm T}_{\rm eff} & 6000\,{\rm K}  \\
\log g  & 0.5\ {\rm(cgs)} \\
\xi_{\rm t} & 3.5\ {\rm km\,s}^{-1} \\
{\rm [M/H]} & -2.5 \\
\end{array}
$
}\\
\hline
  ion & N &{\rule[0mm]{0mm}{4mm}$\overline{W_{\lambda}}$}&$\log\epsilon$&$\sigma_{\rm ltl}$&[el/H]& sun & T$_{\rm cond}$\\
\hline
C\,{\sc i}  &  13 &  63 &  8.00 &  0.13 & $-$0.57 & 8.57 &    40\\
N\,{\sc i}  &   9 &  46 &  8.04 &  0.16 &    0.05 & 7.99 &   123\\
O\,{\sc i}  &   3 &  26 &  8.53 &  0.10 & $-$0.33 & 8.86 &      \\
$[$O\,{\sc i}$]$&2& 102 &  9.16 &  0.08 &    0.30 & 8.86 &      \\
Na\,{\sc i} &   4 &  52 &  5.80 &  0.14 & $-$0.53 & 6.33 &   953\\
Mg\,{\sc i} &   2 &  38 &  5.29 &  0.01 & $-$2.25 & 7.54 &  1327\\
S\,{\sc i}  &   3 &  37 &  6.99 &  0.02 & $-$0.34 & 7.33 &   655\\
Ca\,{\sc i} &   3 &  25 &  4.24 &  0.15 & $-$2.12 & 6.36 &  1505\\
Sc\,{\sc ii}&   1 &  17 &  0.12 &       & $-$3.05 & 3.17 &  1647\\
Ti\,{\sc ii}&   4 &  22 &  2.18 &  0.26 & $-$2.84 & 5.02 &  1573\\
Cr\,{\sc i} &   2 &  49 &  3.26 &  0.04 & $-$2.41 & 5.67 &  1291\\
Cr\,{\sc ii}&   3 &  39 &  3.40 &  0.07 & $-$2.27 & 5.67 &  1291\\
Mn\,{\sc i} &   3 &  28 &  3.60 &  0.13 & $-$1.79 & 5.39 &  1150\\
Fe\,{\sc i} &  35 &  30 &  5.08 &  0.25 & $-$2.43 & 7.51 &  1328\\
Fe\,{\sc ii}&   7 &  52 &  4.87 &  0.19 & $-$2.64 & 7.51 &  1328\\
Zn\,{\sc i} &   4 &  87 &  4.26 &  0.06 & $-$0.34 & 4.60 &   723\\
Y\,{\sc ii} &   1 &  24 &$-$0.26&       & $-$2.50 & 2.24 &  1647\\
Ba\,{\sc ii}&   3 & 105 &  0.13 &  0.15 & $-$2.00 & 2.13 &  1447\\
\hline
\end{tabular}
\end{center}
\end{table}

\section{Conclusion}
In this letter, we have found evidence that the photospheric depletion phenomenon,
which is frequently observed in field RV Tauri stars, also occurs in the
extragalactic members of this pulsation class. The effect on the observed
metallicity of MACHO\,82.8405.15 is very significant ([Fe/H] = $-$2.6), while
the initial iron content was probably only slightly subsolar, as expected in the
LMC. While the degree
of depletion of MACHO\,82.8405.15 is beyond any doubt, the analysis itself can
be improved by adopting a more realistic model atmosphere that accounts for the
observed dynamics and the non-LTE effects that we encountered
in our analysis. Additional spectra at different pulsation phases may reveal
interesting results, both those from the LTE approach of a dynamic photosphere
and from the abundance determination of lines and species that become
accessible at different temperatures.

Although detailed scenario for the depletion process is still lacking, there
is growing evidence \citep{vanwinckel03,deruyter06} that binarity plays a key
role in the creation of a stable dusty circumstellar
environment that is probably needed for the process to occur. High-resolution
infrared data of MACHO\,82.8405.15 would yield invaluable information on the
physical conditions and mineralogy of the circumstellar environment.

Our first extragalactic detection of a depleted photosphere in a luminous
star shows that the depletion process may be rather common elsewhere.
MACHO\,82.8405.15 is a member of a larger sample of luminous LMC RV Tauri
objects. Our recent analysis of one of the other objects
\citep[MACHO\,47.2496.8,][]{reyniers06}
shows, however, a very different chemical result with a
very strong s-process-enhanced photosphere in an intrinsically metal-poor
environment of [Fe/H]\,$=$\,$-$1.4. Clearly the LMC RV\,Tauri stars are also
very diverse chemically. A global study of these objects will enable us to
better constrain both the depletion process itself, as well as the true
evolutionary status of RV\,Tauri stars.

\acknowledgements{It is a pleasure to thank Clio Gielen for the construction
of the SED, Sophie Saesen for the Euler observations, the anonymous referee for
the quick and constructive report, and the Geneva staff for observation time on
the Euler telescope. MR acknowledges financial support from the Fund for
Scientific Research - Flanders (Belgium).}

\bibliographystyle{aa}

\begin{thebibliography}{24}
\expandafter\ifx\csname natexlab\endcsname\relax\def\natexlab#1{#1}\fi

\bibitem[{{Alcock} {et~al.}(1998){Alcock}, {Allsman}, {Alves}, {Axelrod},
  {Becker}, {Bennett}, {Cook}, {Freeman}, {Griest}, {Lawson}, {Lehner},
  {Marshall}, {Minniti}, {Peterson}, {Pollard}, {Pratt}, {Quinn}, {Rodgers},
  {Sutherland}, {Tomaney}, \& {Welch}}]{alcock98}
{Alcock}, C., {Allsman}, R.~A., {Alves}, D.~R., {et~al.} 1998, \aj, 115, 1921

\bibitem[{{Barth{\`e}s} {et~al.}(2000){Barth{\`e}s}, {L{\`e}bre}, {Gillet}, \&
  {Mauron}}]{barthes00}
{Barth{\`e}s}, D., {L{\`e}bre}, A., {Gillet}, D., \& {Mauron}, N. 2000, \aap,
  359, 168

\bibitem[{{De Ruyter} {et~al.}(2006){De Ruyter}, {Van Winckel}, {Maas}, {Lloyd
  Evans}, {Waters}, \& {Dejonghe}}]{deruyter06}
{De Ruyter}, S., {Van Winckel}, H., {Maas}, T., {et~al.} 2006, \aap, 448, 641

\bibitem[{{Deroo} {et~al.}(2005){Deroo}, {Reyniers}, {Van Winckel}, {Goriely},
  \& {Siess}}]{deroo05}
{Deroo}, P., {Reyniers}, M., {Van Winckel}, H., {Goriely}, S., \& {Siess}, L.
  2005, \aap, 438, 987

\bibitem[{{Fokin} {et~al.}(2001){Fokin}, {L{\`e}bre}, {Le Coroller}, \&
  {Gillet}}]{fokin01}
{Fokin}, A.~B., {L{\`e}bre}, A., {Le Coroller}, H., \& {Gillet}, D. 2001, \aap,
  378, 546

\bibitem[{{Gillet} {et~al.}(1990){Gillet}, {Burki}, \& {Duquennoy}}]{gillet90}
{Gillet}, D., {Burki}, G., \& {Duquennoy}, A. 1990, \aap, 237, 159

\bibitem[{{Gillet} {et~al.}(1989){Gillet}, {Duquennoy}, {Bouchet}, \&
  {Gouiffes}}]{gillet89}
{Gillet}, D., {Duquennoy}, A., {Bouchet}, P., \& {Gouiffes}, C. 1989, \aap,
  215, 316

\bibitem[{{Giridhar} {et~al.}(2000){Giridhar}, {Lambert}, \&
  {Gonzalez}}]{giridhar00}
{Giridhar}, S., {Lambert}, D.~L., \& {Gonzalez}, G. 2000, \apj, 531, 521

\bibitem[{{Giridhar} {et~al.}(2005){Giridhar}, {Lambert}, {Reddy}, {Gonzalez},
  \& {Yong}}]{giridhar05}
{Giridhar}, S., {Lambert}, D.~L., {Reddy}, B.~E., {Gonzalez}, G., \& {Yong}, D.
  2005, \apj, 627, 432

\bibitem[{{Jura}(1986)}]{jura86}
{Jura}, M. 1986, \apj, 309, 732

\bibitem[{{L{\`e}bre} {et~al.}(1996){L{\`e}bre}, {Mauron}, {Gillet}, \&
  {Barth{\`e}s}}]{lebre96}
{L{\`e}bre}, A., {Mauron}, N., {Gillet}, D., \& {Barth{\`e}s}, D. 1996, \aap,
  310, 923

\bibitem[{{Lloyd Evans} \& {Pollard}(2004)}]{lloydevans04}
{Lloyd Evans}, T. \& {Pollard}, K.~R. 2004, in ASP Conf. Ser. 310: IAU Colloq.
  193: Variable Stars in the Local Group, ed. D.~W. {Kurtz} \& K.~R. {Pollard},
  344

\bibitem[{{Lodders}(2003)}]{lodders03}
{Lodders}, K. 2003, \apj, 591, 1220

\bibitem[{{Maas} {et~al.}(2005){Maas}, {Van Winckel}, \& {Lloyd
  Evans}}]{maas05}
{Maas}, T., {Van Winckel}, H., \& {Lloyd Evans}, T. 2005, \aap, 429, 297

\bibitem[{{Preston} {et~al.}(1963){Preston}, {Krzeminski}, {Smak}, \&
  {Williams}}]{preston63}
{Preston}, G.~W., {Krzeminski}, W., {Smak}, J., \& {Williams}, J.~A. 1963,
  \apj, 137, 401

\bibitem[{{Reyniers} {et~al.}(2006){Reyniers}, {Abia}, {Van Winckel}, {Lloyd
  Evans}, {Decin}, {Eriksson}, \& {Pollard}}]{reyniers06}
{Reyniers}, M., {Abia}, C., {Van Winckel}, H., {et~al.} 2006, astro-ph/0610240

\bibitem[{{Seares} \& Haynes(1908)}]{seares08}
{Seares}, F.~H. \& Haynes, E.~S. 1908, Laws Observatory Bulletin, University of
  Missouri, 14, 215

\bibitem[{{Shchukina} {et~al.}(2005){Shchukina}, {Bueno}, \&
  {Asplund}}]{shchukina05}
{Shchukina}, N.~G., {Bueno}, J.~T., \& {Asplund}, M. 2005, \apj, 618, 939

\bibitem[{{Th{\' e}venin}(1989)}]{thevenin89}
{Th{\' e}venin}, F. 1989, \aaps, 77, 137

\bibitem[{{Th{\' e}venin}(1990)}]{thevenin90}
---. 1990, \aaps, 82, 179

\bibitem[{{Van Winckel}(2003)}]{vanwinckel03}
{Van Winckel}, H. 2003, \araa, 41, 391

\bibitem[{{Van Winckel} \& {Reyniers}(2000)}]{vanwinckel00}
{Van Winckel}, H. \& {Reyniers}, M. 2000, \aap, 354, 135

\bibitem[{{Van Winckel} {et~al.}(1998){Van Winckel}, {Waelkens}, {Waters},
  {Molster}, {Udry}, \& {Bakker}}]{vanwinckel98}
{Van Winckel}, H., {Waelkens}, C., {Waters}, L.~B.~F.~M., {et~al.} 1998, \aap,
  336, L17

\bibitem[{{Waters} {et~al.}(1992){Waters}, {Trams}, \& {Waelkens}}]{waters92}
{Waters}, L.~B.~F.~M., {Trams}, N.~R., \& {Waelkens}, C. 1992, \aap, 262, L37

\end{thebibliography}



\end{document}